\documentstyle[aps,12pt]{revtex}

\begin{document}
\title{CHIRAL QUARK PERSPECTIVE OF\\
THE PROTON SPIN AND FLAVOR PUZZLES\thanks{
CMU-HEP95-16, DOE-ER/40682-105, hep-ph 9510\#\#\#.}}
\author{{\sc T. P. Cheng}}
\address{Department of Physics, University of Missouri, St Louis, MO 63121}
\author{{\sc Ling-Fong Li}}
\address{Department of Physics, Carnegie Mellon University, Pittsburgh, PA
15213\\
\smallskip\ }
\maketitle

\begin{abstract}
{\bf Abstract: }The chiral quark model wih a broken-U(3) symmetry gives a
simple and unified account for the various proton spin and flavor puzzles,
as well as the octet baryon magnetic moments.
\end{abstract}

\section{The Proton Spin \& Flavor Puzzles}

Ever since the 1960's it has been known that the simple nonrelativistic
quark model gives a good approximate description of low energy hadron
physics. In particular, the simple quark model (sQM) can give a good account
of the baryonic spectroscopy and magnetic moments. The proton is pictured to
be composed of three almost-free constituent quarks confined within a
distance on the order of a fermi. There is no quark sea in the sQM.

However, in recent years experimental findings, by EMC, SMC, E142, E143\cite
{emc}, NMC and NA51\cite{nmc} have been interpreted as indicating that the
proton has a spin and flavor structure that deviates significantly from the
sQM expectations. Namely, the effects associated with the quark sea have
been found to be not negligible. For example, in sQM the proton spin comes
simply from the addition of its constituent quark spins. For each $q$-flavor
quark contribution to the proton spin $\Delta q=\left( q_{\uparrow
}-q_{\downarrow }\right) +\left( \overline{q}_{\uparrow }-\overline{q}%
_{\downarrow }\right) \equiv \Delta _q+\Delta _{\overline{q}}$, we have
\begin{equation}
\Delta u=\frac 43,\;\Delta d=-\frac 13,\;\Delta s=0,\;\Delta \Sigma =1,
\label{sqmspin}
\end{equation}
$\Delta \Sigma $ being the sum. An analysis using octet baryon weak axial
charges and the polarized lepton-nucleon DIS data\cite{emc} has shown that
the Ellis-Jaffe sum rule\cite{ejsr} is violated, and it gives the spin
components\cite{ellisk} :
\begin{equation}
\Delta u=0.83,\,\Delta d=-0.42,\,\Delta s=-0.10,\,\Delta \Sigma =0.31
\label{emcspin}
\end{equation}
with an estimated error of $0.06$ for each flavor's contribution. This
discrepancy is puzzling in view of the fact that the same sQM spin structure
(\ref{sqmspin}) leads to a fairly good description of the octet baryon
magnetic moments. We note that each $\left( \Delta q\right) _{\text{exptl}}$
in (\ref{emcspin}) is more negative than the corresponding $\left( \Delta
q\right) _{\text{sQM}}$ in (\ref{sqmspin}). This means that the quark sea
must be polarized strongly in the opposite direction to the proton spin.

The magnetic moment of a baryon is related to the quark and antiquark
polarizations as

\begin{eqnarray}
\mu _B &=&\sum_{q=u,d,s}\left[ \left( \Delta _q\right) _B\mu _q+\left(
\Delta _{\overline{q}}\right) _B\mu _{\overline{q}}\right]  \label{mm} \\
\ &=&\sum \left[ \left( \Delta _q\right) _B-\left( \Delta _{\overline{q}%
}\right) _B\right] \mu _q\equiv \sum \left( \widetilde{\Delta q}\right)
_B\mu _q.  \nonumber
\end{eqnarray}
Using flavor-$SU(3)$ one can relate all $\left( \widetilde{\Delta q}\right)
_B$ to proton's $\widetilde{\Delta q}$. An analysis using both the octet
baryon $\mu _B^{\prime }s,$ which are related to the {\em difference} of the
quark and antiquark polarizations inside the proton, and the proton spin's
quark components (\ref{emcspin}), to the {\em sum,} shows that antiquark
polarizations inside the proton $\Delta _{\overline{q}}$ is small\cite
{clphlett}.

The NMC measurement of the muon scatterings off proton and neutron target
shows that the Gottfried sum rule is violated\cite{nmc}. This has been
interpreted as showing a proton quark sea being not symmetric with respect
to the $u$ and $d$ quark pairs: $\overline{d}>\overline{u}.$ The conclusion
has been confirmed by the asymmetry measurement (by NA51) in the Drell-Yan
process with proton and neutron targets, which yield $\overline{d}\simeq 2%
\overline{u}$ at the quark momentum $x=0.18.$ These results contradict our
expectation of $\overline{d}\simeq \overline{u}:$ since $u$-, $d$-quarks are
similar in mass and the quark sea should be created by the
flavor-independent gluon emissions. In fact there had long been some
indication that the flavor content of the proton quark sea may not be as
simple as one would expect. The size of the pion-nucleon sigma term\cite
{sigterm} of $45\,MeV$ means that the OZI rule for the strange quark is
strongly violated, and this can be translated into a statement that the
fraction of strange quarks in the proton, averaged over all momenta, is not
small, $f_s\simeq 0.18.$

\section{Proton Spin \& Flavor Contents in the Chiral Quark Model}

The basic idea of chiral quark model\cite{mgtheor} is that the energy scale
associated with chiral symmetry breaking is much larger than the QCD
confinement scale. Thus in the interior of a hadron (but not so short a
distance when perturbative QCD becomes operative) the relevant degrees of
freedom are the {\em quasiparticles} of quarks, gluons, and the Goldstone
bosons of chiral symmetry. Here, the quarks propagate in a ground state
filled with $\overline{q}q$ condensates and gain in mass giving a
constituent quark mass around a third of the nucleon mass. The quark-gluon
interactions of the underlying QCD bring about the chiral symmetry breaking
and Goldstone excitations. But, when the description is organized in terms
of the quasiparticle effective fields, the remanent gluon coupling is
expected to be small. Thus the most important interaction in this regime is
the coupling among the internal Goldstone bosons and quarks. In Ref.\cite
{ehq} and \cite{clprl} it has been suggested that such interactions can
yield a simple and natural explanation of the spin and flavor puzzles.

A quark sea created through internal Goldstone boson (GB) emissions by a
valence quark,
\begin{equation}
q_{\uparrow }\rightarrow GB+q_{\downarrow }^{\prime }\rightarrow \left( q\;%
\overline{q^{\prime }}\right) _0\;q_{\downarrow }^{\prime }  \label{gbemit}
\end{equation}
has just the desired spin polarization features. The coupling of the
pseudoscalar Goldstone boson to the quarks will flip the polarization of the
quark: $q_{\uparrow }$ $\rightarrow $ $q_{\downarrow }^{\prime }.$ We note
that the final state $q_{\downarrow }^{\prime }$ carries {\em all} the
polarization of the quark-sea, as the pair $\left( q\;\overline{q^{\prime }}%
\right) _0$ --- coming out of the Goldstone boson --- must be in the
spin-zero combination:
\begin{equation}
\left( q\;\overline{q^{\prime }}\right) _0=\frac 1{\sqrt{2}}\left(
q_{\uparrow }\overline{q_{\downarrow }^{\prime }}-q_{\downarrow }\overline{%
q_{\uparrow }^{\prime }}\right) .  \label{spinzero}
\end{equation}
In this manner, the quark sea adds a {\em negative} amount to each of the $%
\Delta q^{\prime }s$ in (\ref{sqmspin}), and, from (\ref{spinzero}) we also
have the ''no antiquark polarization'' feature of $\Delta _{\overline{q}}=0,$
thus $\Delta q=\widetilde{\Delta q},$ as required by the phenomenological
analysis discussed above.

The GB emissions create a quark sea having just the right flavor structures.
We have the processes $u\rightarrow \pi ^{+}d\rightarrow u\overline{d}d,$ $%
u\rightarrow K^{+}s\rightarrow u\overline{s}s$, but not $u\rightarrow \pi
^{-}...\rightarrow \overline{u}d...,$ because there is no charge $5/3$
quarks. Even though this flavor asymmetry may be diluted somewhat by the
emission of $\pi ^o,\;\eta \;$and $\eta ^{\prime }$ GB modes, the valence $u$
is favored to produce $\overline{d}d$ and $\overline{s}s$, while $d$ is
favored to produce $\overline{u}u$ and $\overline{s}s.$ Since proton has two
valence $u$ quarks and one valence $d$, this GB emission mechanism can
easily produce a quark sea with more $d$-pairs than $u$-pairs, and also more
strange quarks if their emissions had not been suppressed by heavier strange
GB's.

We have advocated a chiral quark model with a broken $U(3)=SU(3)\times U(1)$
symmetry\cite{clprl}. In this version there are two parameters which
correspond to the octet GB and singlet GB couplings to quarks, $g_8$ and $%
g_1 $. A choice of $a=0.1$ as the probability $\propto \left| g_8\right| ^2$
for the $u$ quark to emit a $\pi ^{+}$ (and its $SU(3)$ generalizations),
and coupling ratio $\varsigma \equiv g_1/g_8=-1.2$ has been found to give a
good account for all the observed proton's spin and flavor structures, as
well as the octet baryon $\mu _B^{\prime }s$\cite{clphlett}$.\;$(See Table
1, all $\mu ^{\prime }$s are in nucleon magnetons.) In fitting the $\mu
_B^{\prime }s $, we have constrained the quark moments as $\mu _u=-2\mu
_d,\,\mu _s/\mu _d=0.6$, and have adjusted the remaining independent value
of $\mu _u$ to get a good fit.\

\section{Discussion}

One should keep in mind that our result is deduced basically from an $SU(3)$
symmetric calculation. The only $SU(3)\;$breaking effect that has been taken
into account is the different moments $\mu _s/\mu _d=0.6$ reflecting the
different constituent quark masses of $m_{u,d}$ and $m_s.$ Thus we do not
really expect a better than 20 to 30\% agreements from the model predictions.

It is gratifying that an elementary calculation in a physically
well-motivated model can, in a simple and unified way, account for the
proton spin and flavor puzzles. Clearly, one needs to incorporate the $SU(3)$
breaking effects more systematically. To do this, and to find out the $x$
and $Q^2$ dependences of the quark number and spin densities, one must know
more about the GB modes propagating in the interior of the hadron.
Nevertheless, the success of the chiral quark model calculations seem to
indicate that the original constituent quark model is generally correct in
its description of the low energy hadron physics. It only needs to be
augmented by a quark sea which is perturbatively generated by the valence
quarks through internal GB emissions.

\begin{center}
\begin{tabular}{ccc}
\hline\hline
& Experimental & Chiral quark model \\
& value & $a=0.1,\;\varsigma =-1.2$ \\ \hline
$\overline{d}-\overline{u}$ & $\;0.147\pm 0.026$ & $\;\;0.147$ \\
$\overline{u}/\overline{d}$ & $\;0.51\pm 0.09$ & $\;0.53$ \\
$f_s$ & $\;0.18\pm 0.03$ & $\;0.19$ \\
$\Delta u$ & $\;0.83\pm 0.05$ & $\;0.79$ \\
$\Delta d$ & $-0.42\pm 0.05$ & $-0.32$ \\
$\Delta s$ & $-0.10\pm 0.05$ & $-0.10$ \\
$\Delta \Sigma $ & $\;0.31\pm 0.05$ & $\;0.37$ \\ \hline
$\mu _p$ & $\;2.79\pm 0.00$ & $\;2.69$ \\
$\mu _n$ & $-1.91\pm 0.00$ & $-1.88$ \\
$\mu _{\Sigma ^{+}}$ & $\;2.48\pm 0.05$ & $\;2.56$ \\
$\mu _{\Sigma ^{-}}$ & $-1.16\pm 0.03$ & $-1.10$ \\
$\mu _{\Xi ^0}$ & $-1.25\pm 0.03$ & $-1.37$ \\
$\mu _{\Xi ^{-}}$ & $-0.68\pm 0.03$ & $-0.48$ \\
$\mu _\Lambda $ & $-0.61\pm 0.01$ & $-0.60$ \\
$\mu _{\Lambda \Sigma }$ & $-1.60\pm 0.08$ & $-1.58$ \\ \hline\hline
\end{tabular}

TABLE 1
\end{center}

This work is supported at UM-St Louis by the National Science Foundation
(PHY-9207026), and at CMU by the US Department of Energy
(DE-FG02-91ER-40682).

\end{document}